------------------------------------------------------------------------
\documentstyle[12pt,preprint,aps]{revtex}

\begin{document}

\draft

\title{Low-energy three-body charge transfer reactions with Coulomb
interaction in the final state\thanks{To appear in Journal of Physics B}}

\author{Renat A. Sultanov and Sadhan K. Adhikari}

\address{Instituto de F\'\i sica Te\'orica,
Universidade Estadual Paulista,\\
01405-900 S\~{a}o Paulo, S\~{a}o Paulo, Brazil}

\maketitle

\vspace{2cm}

\begin{abstract}

Three-body charge transfer reactions with Coulomb interaction in the final
state are considered in the framework of coordinate-space
integro-differential Faddeev-Hahn-type equations within two- and six-state
close coupling approximations.  The method is employed to study direct
muon transfer in low-energy collisions of the muonic hydrogen H$_\mu$ by
helium (He$^{++}$) and lithium (Li$^{+++}$) nuclei.  The experimentally
observed isotopic dependence is reproduced.  

\end{abstract}

%
%

\narrowtext
\section{INTRODUCTION}
\label{sec:intro}

Experimental investigations of the low-energy muon-transfer reactions in
collisions of muonic hydrogen H$_\mu$ (bound state of a hydrogen isotope
and muon $\mu^-$) with nuclei of 
charge $Z_1>1$ are of
importance for  muon catalyzed fusion cycle \cite{Rafelski-1991}.
The study of such  collisions  involving three charged particles is also
very interesting
from a theoretical point of view as an example of rearrangement scattering
with Coulomb interaction in the final state. Such reactions with
post-collision Coulomb interaction between clusters 
appear frequently in atomic and molecular physics
\cite{Post-1999}. In the following we
develop a general formalism for dealing with such reactions and as an
example apply it to study some muon-transfer processes.

Recently, there has been considerable
experimental interest 
in the study of the muon-transfer  reaction
in collision of the muonic atoms
with He$^{++}$ \cite{Jacot-1998,Tresch1-1998} and also
with charges $Z_1>3$
\cite{Wert-1998,iso1,Thalmann-1998,Mulhauser-1993,Schell-1993},
e.g. oxygen (O$^{8}$),
neon (Ne$^{10}$), argon (Ar$^{18}$) etc. It was found that
contrary to the smooth
$Z$-dependence expected from the semiclassical Landau-Zener formula
\cite{Landau-1932}
the experimental muon
transfer rates for reactions like
\begin{equation}
(\mbox{H} _\mu)_{1s} + {\mbox{X}^Z} \rightarrow 
\mbox{X}^Z_\mu
+ \mbox{H} \label{1}  
\end{equation}
depend in a complicated manner on the charge $Z$ \cite{Schell-1993}. 
Here $\mbox{H}$ stands for the hydrogen isotopes  ${^1{\mbox H}}$ or
${^2{\mbox H}}$ and $\mbox{X}^Z$  stands for the target nuclei.
Another phenomenon which has not yet found a satisfactory theoretical
explanation is the measured isotope effect, e.g.
the trend  of the direct transition rates of reactions (\ref{1})
for X$^Z=$ O$^{8}$ \cite{Mulhauser-1993},
Ne$^{10}$ \cite{Schell-1993},  Ar$^{18}$ \cite{iso1},
and Xe$^{54}$ \cite{iso2}.
In cases of O$^{8}$, Ar$^{18}$ and Xe$^{54}$
the direct transfer rate decreases  with increasing
the mass of the H isotope.
Theoretical analyses \cite{iso3} also support this trend.
The experimental  results for Ne$^{10}$ \cite{Schell-1993}
and sulphur dioxide \cite{Mulhauser-1993}
differ considerably from the theoretical predictions.
Moreover, several experiments performed in recent
years have put into evidence the complex structure of the time
distributions of the X-rays following transfer from muonic hydrogen
isotopes to heavier elements \cite{Schell-1996}.

The proper theoretical analysis of charge transfer reaction (\ref{1}) 
becomes extremely complicated numerically as the charge $Z$ increases
because of the presence of the strong Coulomb interaction in the final
state. Traditionally, in theoretical studies, such Coulombic systems with
two heavy (nuclei) and one light (muon)  particles are considered within
the framework of the two-state molecular Born-Oppenheimer approximation
\cite{Matv-1973,W-1992}.  In another study, a semiclassical model based on
Faddeev-type scattering equations has been used \cite{Sultanov-1999}. It
would be of interest to perform a full quantum mechanical consideration in
view of the fact that the muon is not so light compared to the nucleon and
compare with the approximate calculations mentioned above.

Here we develop   a quantum mechanical approach based on Faddeev-Hahn-type
equations for a careful reinvestigation of these three-body direct 
charge-transfer reactions with strong Coulomb repulsion in the final
state.  As a first step towards a model  solution of
this complicated problem, we apply  this  detailed few-body method to
the study of direct muon-transfer reaction (\ref{1})
for $X^Z = $
$^3{\mbox {He}}^{++}$, 
$^4{\mbox {He}}^{++}$, $^6{\mbox {Li}}^{+++}$ and $^7{\mbox{Li}}^{+++}$.
This study with lighter nuclei is expected to lead to faster numerical
convergence than the heavier targets.  However,
our approach   is equally applicable for heavier targets with
higher charges, although the convergence could be slow in these cases.
These   studies with heavier targets would be  interesting future works.

For the three-charged-particle system, say 
$(^7{\mbox {Li}}\ {^2{\mbox H}}\ \mu)$, only two
asymptotic two-cluster configurations are possible, i.e. 
$(^2{\mbox H }_\mu)-{^7{\mbox {Li}}}$
and
$(^7{\mbox {Li}}_\mu)-{^2{\mbox H}}$.  
For the theoretical treatment of such a  three-body rearrangement process,
Faddeev-type equations \cite{Faddeev-1961}, especially the modified
version proposed by Hahn \cite{Hahn-1968}, appear to be
very suitable. 
The  two possible 
asymptotic configurations of the above rearrangement problem  are
conveniently tackled by a set
of two coupled Faddeev-Hahn-type
equations for components $\Psi_1$ and $\Psi_2$ of the wave function $\Psi
= \Psi_1 + \Psi_2$, where each component carrys the
asymptotic boundary condition for a specific configuration
\cite{renat,renat98}.
These equations are very useful to
incorporate distortion potentials for specific initial and final 
asymptotic states \cite{Hahn-1972}.
It is
possible to include the final-state Coulomb interaction explicitly in
these
equations, so that a low-order approximation to these equations produces 
the correct asymptotic behavior \cite{Hahn-1972}.


We solve the integro-differential form of the Faddeev-Hahn equation by
the close-coupling approximation scheme involving up to six states. This
procedure consists in expanding the wave function components $\Psi_1$ and
$\Psi_2$ in terms of eigenfunctions of subsystem Hamiltonians in initial
and final channels, respectively.  Although, these subsystem
eigenfunctions are not orthogonal to each other, the components $\Psi_1$
and $\Psi_2$ satisfy a coupled set of equations incorporating the correct
asymptotic behavior of the wave function. Consequently, there is no
problem of overcompleteness as encountered in similar expansion approaches
for rearrangement reactions based on the Schr\"odinger equation. The
resultant coupled Faddeev-Hahn-type equations are then projected on the
expansion functions. After a partial-wave projection this leads to a set
of one-dimensional coupled integro-differential equations for the
expansion coefficients, which is solved numerically.

In Sec. II  we develop the formalism. We have calculated transfer rates
for reaction (1) for 
$\mbox{H} = {^1{\mbox H}}$ or
${^2{\mbox H}}$ and
$\mbox{X}^Z$  =  $^3{\mbox {He}}^{++}$,
$^4{\mbox {He}}^{++}$, 
$^6{\mbox {Li}}^{+++}$ or
$^7{\mbox {Li}}^{+++}$ using a two-state close-coupling approximation,
and for
$\mbox{H} =  {^2{\mbox H}}$ and 
$\mbox{X}^Z = ^3{\mbox {He}}^{++}$, 
$^6{\mbox {Li}}^{+++}$ or $^7{\mbox {Li}}^{+++}$
using six-state close-coupling approximations.
Our results
obtained for muon-transfer rates  from hydrogen to helium
and lithium are given in
Sec. III and compared  with those of other investigations.
We also present a summary and outlook in the concluding part of
this section.

\narrowtext
\section{Theoretical Formulation}

Let us take the system of units to be $e=\hbar=m_\mu=1$, where $m_\mu$
($e$) is
the muonic  mass  (charge), and
denote, the heavy nuclei 
($^3{\mbox {He}}$, $^4{\mbox {He}}$, 
$^6{\mbox {Li}}$, etc.)  by ${\sf 1}$, the hydrogen isotopes
($^1{\mbox H}$, $^2{\mbox H}$ or $^3{\mbox H}$)
by ${\sf 2}$ and muon by ${\sf 3}$.
Below the three-body breakup threshold, following 
two-cluster asymptotic configurations
are possible in the system {\sf 123}:  $({\sf 23})\ -\ {\sf 1}$ and
$({\sf 13})\ -\ {\sf 2}$. These two configurations 
  correspond to two distinct physical channels, also denoted by  
1 and 2. 
These configurations  
are 
 determined by the Jacobi coordinates
$(\vec r_{j3}, \vec \rho_k)$
\begin{equation}
\vec r_{j3} = \vec r_3 - \vec r_j,
\hspace{6mm} \vec \rho_k =
(\vec r_3 + m_j\vec r_j) / (1 + m_j) - \vec r_k,
\hspace{6mm} j\not=k=1, 2,
\label{eq:coord}
\end{equation}
$\vec r_{i}$, $m_{i}$ are coordinates and
masses of the particles $i=1, 2, 3,$ respectively. 

Let us introduce
the total three-body wave function as a sum of two components
\begin{equation}
\Psi(\vec r_1, \vec r_2, \vec r_3) \ =\  \Psi_1 (\vec r_{23},\vec \rho_1)
\ + \ \Psi_2 (\vec r_{13},\vec \rho_2),
\label{eq:total2}
\end{equation}
where $\Psi_1 (\vec r_{23},\vec \rho_1)$
is quadratically integrable over the variable
$\vec r_{23}$, and  $\Psi_2 (\vec r_{13},\vec \rho_2)$ over the
variable $\vec r_{13}$. The components $\Psi_1$ and $\Psi_2$ carry the
asymptotic
boundary condition for channels 1 and 2, respectively.
The second component is responsible for
pure Coulomb interaction in the final state.
These components satisfy the following 
 set of two coupled equations
\begin{eqnarray}
\begin{array}{l}
(E - H_0 - V_{23})\Psi_1 (\vec r_{23}, \vec \rho_1) =
(V_{23} + V_{12} - U_C)
\Psi_2 (\vec r_{13}, \vec \rho_2)
\vspace{7mm} \\
(E - H_0 - V_{13} - U_C)
\Psi_2 (\vec r_{13}, \vec \rho_2) =
(V_{13} + V_{12})\Psi_1 (\vec r_{23}, \vec \rho_1)\;,
\end{array}
\label{eq:fadd5}
\end{eqnarray}
where $E$ is the center-of-mass energy, $H_0$ is the total kinetic energy
operator, and $V_{ij} (r_{ij})$
are pair-interaction potentials $(i \not= j = 1, 2, 3)$,
and $U_C$ is a distortion interaction, e.g. Coulomb repulsion
in the final state between clusters $(^3{\mbox {He}},\mu)$
and $ {^2{\mbox H}}$ in the 
case of $^3{\mbox {He}}\ {^2{\mbox H}}\ \mu$ system 
\begin{equation}
U_C = \frac{(Z_1 - 1)Z_2}{\rho_2}.
\end{equation}
Here $Z_1$ is the charge of $^3{\mbox {He}}$ and $Z_2 (=1)$ is the
charge of the
hydrogen  isotope. 
By adding the two 
equations (\ref{eq:fadd5}) we find that they are equivalent to 
the Schr\"odinger
equation.
  For  energies below the three-body
breakup threshold they possess the same advantages as the Faddeev
equations, since they are formulated for
the wave function components 
with 
correct physical
asymptotic behavior.

The component $\Psi_1$ carries the asymptotic
behavior in elastic and inelastic channels:
\begin{equation}
\Psi_1(\vec r_{23}, \vec \rho_1)\ 
\mathop{\mbox{\large$\sim$}}\limits_{\rho_1 \rightarrow + \infty}\ 
e^{ik^{(1)}_1z}\varphi_1(\vec r_{23})\ + \ 
\sum_n A_n^{\mbox{\scriptsize{el/in}}}(\Omega_{\rho_1})
e^{ik^{(1)}_n\rho_1}\varphi_n(\vec r_{23})/\rho_1\; .
\end{equation}
The component $\Psi_2$ carries the Coulomb asymptotic
behavior in the transfer channels:
\begin{equation}
\Psi_2(\vec r_{13}, \vec \rho_{2})
\mathop{\mbox{\large$\sim$}}\limits_{\rho_2 \rightarrow + \infty}
\sum_{ml} A_{ml}^{\mbox{\scriptsize{tr}}}(\Omega_{\rho_2})
e^{i(k^{(2)}_m
\rho_2 - \pi l /2 + \tau_l - \eta / 2 k^{(2)}_m \ln2k^{(2)}_m
\rho_2)}
\varphi_m(\vec r_{13})/\rho_2,
\label{eq:tr}
\end{equation}
where $e^{ik^{(1)}_1z} \varphi_1(\vec r_{23})$ is the incident wave,
$\varphi_n(\vec r_{j3})$ the $n$-th excited bound-state wave function
of pair $(j3)$, $k_n^{(i)} = \sqrt{2M_{i}(E- E_n^{(j)})}$,
with $M_i^{-1}= m_i^{-1} + (1 + m_j)^{-1}\ $. Here
$E_n^{(j)}$ is the binding energy of  $(j3)$, $i\ne j=1,2$, 
$A^{\mbox{\scriptsize{el/in}}}(\Omega_{\rho_{1}})$
and $A^{\mbox{\scriptsize{tr}}}(\Omega_{\rho_{2}})$
are the scattering amplitudes in  the elastic/inelastic and transfer
channels. The Coulomb parameters in the second transfer channel
are \cite{Mott-1965}
\begin{equation}
\tau_l = \makebox{arg} \Gamma(l+1+i\eta/2k^{(2)}_m)
\hspace{10mm}
\mbox{and}
\hspace{10mm}
\eta = 2M_2(Z_1-1)/k^{(2)}_n.
\end{equation}
This approach simplifies  the
solution procedure and   provides the correct asymptotic
behavior of the solution below the 3-body breakup threshold.

Let us write down (\ref{eq:fadd5})
in terms of the adopted notations
\begin{eqnarray}
\left[ E + \frac{\nabla^2_{\vec \rho_k}}{2 M_k} +
\frac{\nabla^2_{\vec r_{j3}}}{2 \mu_{j}} -
V_{j3} - U_C\delta_{k2}\right]
\Psi_k (\vec r_{j3}, \vec \rho_k)\ = \ 
(V_{j3} + V_{jk}  - U_C\delta_{j2})
\Psi_{j}(\vec r_{k3}, \vec \rho_j)\;,
\label{eq:fadd8}
\end{eqnarray}
here $j\ne k =1, 2$, $M_k^{-1}= m_k^{-1} + (1 + m_j)^{-1}\ $
and $ \mu_j^{-1} = 1 + m_j^{-1}.$
We are using the Jacobi coordinates 
\begin{eqnarray}
\vec \rho_j = \vec r_{j3} - \beta_k \vec r_{k3}, \hspace{4mm}
\vec r_{j3} = \frac{1}{\gamma} (\beta_k\vec \rho_k + \vec \rho_j)
\hspace{4mm}\mbox{and}\hspace{4mm}
\vec r_{jk} = \frac{1}{\gamma} (\sigma_j\vec \rho_j - \sigma_k
\vec \rho_k)\;,
\end{eqnarray}
with 
\begin{equation}
\beta_k = \frac{m_k}{1 + m_k},\hspace{4mm}\sigma_k = 1 - \beta_k
\hspace{4mm}\mbox{and}
\hspace{4mm} \gamma = 1 - \beta_k \beta_j.
\end{equation}
For solving (\ref{eq:fadd8}) we expand the wave function
components in terms of bound states in initial and final channels,
and project this equation on these bound states. The expansion of the wave
function is given by
\begin{equation}
\Psi_k(\vec r_{j3}, \vec \rho_{k}) \approx \sum_{LM\lambda l}
\sum_n \frac{1}{\rho_k}
f_{nl\lambda}^{(k)LM}(\rho_k) R_{nl}^{(k)}(r_{j3})\
\left \{ Y_{\lambda}(\hat \rho_k) \otimes
Y_l(\hat r_{j3}) \right \}_{LM},
\label{eq:expan}
\end{equation}
where $(nl\lambda)\equiv \alpha$ are quantum numbers of a
three-body state and $L$ is the total angular momentum of the
three-body system obtained by coupling $l$ and $\lambda$, $Y_{lm}$'s are
the  spherical harmonics, $R_{nl}^{(k)}(r_{j3})$ the radial part of the
hydrogen-like bound-state
wave function, $f_{nl\lambda}^{(k)LM}(\rho_k)$ are the unknown expansion 
coefficients.
This prescription
is similar to that adopted in the close-coupling approximation.
After a proper angular momentum projection,
 the  set of integro-differential equations
for the unknown expansion functions $f^{(k)}_{\alpha}(\rho_k)$ can be
written as 
\begin{eqnarray}
\left[ (k^{(1)}_n)^2\ +\ \frac{\partial^2}
{\partial \rho_1^2}\ -\ 
\frac{\lambda (\lambda + 1)}{\rho_1^2}
\right] f_{\alpha}^{(1)}(\rho_1) =
g_1\sum_{\alpha'}
\frac{\sqrt{(2\lambda + 1)(2\lambda^{\prime} + 1)}}{2L+1}
\; \nonumber \\
\int_{0}^{\infty} d \rho_{2}
f_{\alpha^\prime}^{(2)}(\rho_{2})\int_{0}^{\pi}
d \omega \sin\omega
R_{nl}^{(1)}(|\vec{r}_{23}|)
\left[-\frac{1}{|\vec {r}_{23}|} + \frac{Z_1}{|\vec {r}_{12}|}
 - U_C\right]
R_{n'l'}^{(2)}(|\vec{r}_{13}|)
\;\nonumber \\
\rho_1 \rho_2 \sum_{mm'} D_{mm'}^L(0, \omega, 0)C_{\lambda 0lm}^{Lm}
C_{\lambda' 0l'm'}^{Lm'}
Y^*_{lm}(\nu_1, \pi) Y_{l'm'}(\nu_{2}, \pi)\;,
\label{eq:most1}
\end{eqnarray}
\begin{eqnarray}
\left[ (k^{(2)}_n)^2\ +\ \frac{\partial^2}
{\partial \rho_2^2}\ -\
\frac{\lambda (\lambda + 1)}{\rho_2^2} - U_C
\right] f_{\alpha}^{(2)}(\rho_2) =
g_2\sum_{\alpha'}
\frac{\sqrt{(2\lambda + 1)(2\lambda^{\prime} + 1)}}{2L+1}
\; \nonumber \\
\int_{0}^{\infty} d \rho_{1}
f_{\alpha^\prime}^{(1)}(\rho_{1})\int_{0}^{\pi}
d \omega \sin\omega
R_{nl}^{(2)}(|\vec{r}_{13}|)
\left[-\frac{Z_1}{|\vec {r}_{13}|} + \frac{Z_1}{|\vec {r}_{12}|}\right]
R_{n'l'}^{(1)}(|\vec{r}_{23}|)
\;\nonumber \\
\rho_{2} \rho_1 \sum_{mm'} D_{mm'}^L(0, \omega, 0)C_{\lambda 0lm}^{Lm}
C_{\lambda' 0l'm'}^{Lm'}
Y^*_{lm}(\nu_2, \pi) Y_{l'm'}(\nu_{1}, \pi)\;.
\label{eq:most2}
\end{eqnarray}
Here $g_k=4\pi M_k/\gamma^{3}$,
$\gamma=1-m_km_{j}/((1+m_k)(1+m_j))$, $\alpha' \equiv (n'l'\lambda')$,
$D_{mm'}^L(0, \omega, 0)$ the Wigner function,
$C_{\lambda 0lm}^{Lm}$ the Clebsh-Gordon coefficient,
$\omega$ is the angle between the Jacobi coordinates
$\vec \rho_i$ and $\vec \rho_{i'}$, $\nu_i$ is the angle between 
$\vec r_{i'3}$ and $\vec \rho_i$, $\nu_{i'}$ is the angle
between $\vec r_{i3}$ and $\vec \rho_{i'}$.
The following relations are  useful for numerical treatment
\begin{eqnarray}
\sin \nu_i = \frac{\rho_{i'}}{\gamma r_{i'3}}\sin\omega
\hspace{5mm}
\mbox{and}
\hspace{5mm}
\cos \nu_i = \frac{1}{\gamma r_{i'3}}
(\beta_i \rho_i + \rho_{i'} \cos \omega)\hspace{5mm}
(i \ne i' = 1,2).
\end{eqnarray}

To find unique solution to
(\ref{eq:most1})$-$(\ref{eq:most2}),
appropriate boundary conditions are to be considered. First we impose
$f_{nl}^{(i)}(0)
\mathop{\mbox{\large$=$}}0$.
For the present scattering  problem with $1 +(23)$ as the initial state,
in the asymptotic region, two solutions to
(\ref{eq:most1})$-$(\ref{eq:most2}) satisfy the following boundary
conditions
\begin{eqnarray}
\left\{
\begin{array}{l}
f_{1s}^{(1)}(\rho_1)
\mathop{\mbox{\large$\sim$}}\limits_{\rho_1 \rightarrow + \infty}
\sin(k^{(1)}_1\rho_1) + {\it K}_{11}\cos(k^{(1)}_1\rho_1)\;,
\vspace{1mm}\\
f_{1s}^{(2)}(\rho_2)
\mathop{\mbox{\large$\sim$}}\limits_{\rho_2 \rightarrow + \infty}
\sqrt{v_1 / v_2}{\it K}_{12}
\cos(k^{(2)}_1\rho_2 - \eta / 2k^{(2)}_1 \ln2k^{(2)}_1\rho_2)\;,\\
\end{array}\right.
\label{eq:cond88}
\end{eqnarray}
where 
$\it K_{ij}$ are the appropriate coefficients.
For scattering
with ${\sf 2} + ({\sf 13})$ as the initial state, we have the following
conditions
\begin{eqnarray}
\left\{
\begin{array}{l}
f_{1s}^{(1)}(\rho_1)
\mathop{\mbox{\large$\sim$}}\limits_{\rho_1 \rightarrow + \infty}
\sqrt{v_2 / v_1}{\it K}_{21}\cos(k^{(1)}_1\rho_1)\;,
\vspace{1mm}\\
f_{1s}^{(2)}(\rho_2)
\mathop{\mbox{\large$\sim$}}\limits_{\rho_2 \rightarrow + \infty}
\sin(k^{(2)}_1\rho_2 - \eta / 2k^{(2)}_1 \ln2k^{(2)}_1\rho_2) +
{\it K}_{22}
\cos(k^{(2)}_1\rho_2 -
\;\nonumber \\
\hspace{26mm}\eta / 2k^{(2)}_1 \ln2k^{(2)}_1\rho_2)\;,\\
\end{array}\right.
\label{eq:cond8888}
\end{eqnarray}
where $v_i$ ($i=1,2$) are velocities in channel $i$. 
In the absence of Coulomb interaction $U_C$ in the final channel,
$\it K_{ij}$ are the components of the on-shell $K$-matrix
\cite{Mott-1965}.
With the following change of variables in
(\ref{eq:most1})$-$(\ref{eq:most2})
\begin{eqnarray}
\begin{array}{l}
{\sf f}_{1s}^{(1)}(\rho_1)=
f_{1s}^{(1)}(\rho_1)-\sin(k^{(1)}_{1}\rho_1),
\vspace{4mm}\\
{\sf f}_{1s}^{(2)}(\rho_2)=
f_{1s}^{(2)}(\rho_2)
-\sin(k^{(2)}_1\rho_2 - \eta / 2k^{(2)}_1 \ln2k^{(2)}_1\rho_2)\;,
\end{array}
\end{eqnarray}
we obtain two sets of inhomogeneous equations which are
solved numerically. The coefficients $\it K_{ij}$ are obtained from the
numerical solution of the Faddeev-Hahn-type equations.
The cross sections are given by
\begin{eqnarray}
\sigma_{ij} =
\frac{4\pi}{k_1^{(i)2}}\frac{\delta_{ij}D^2 + {\it K}_{ij}^2}
{(D - 1)^2 + ({\it K}_{11} + {\it K}_{22})^2},
\end{eqnarray}
where $i,j=1,2$ refer to the two channels and
$D = {\it K}_{11}{\it K}_{22} - {\it K}_{12}{\it K}_{21}$.
When $k_1^{(1)} \rightarrow 0$:
$\sigma_{\mbox{tr}} \equiv  \sigma_{12} \sim 1/k_1^{(1)}$.
For comparison with experimental low-energy data it is very useful
to calculate the transfer rates
\begin{equation}
\lambda_{\mbox{tr}} = \sigma_{\mbox{tr}} v N_0 ,
\end{equation}
with $v$ being the relative velocity of the incident
fragments and $N_0$ the liquid-hydrogen density chosen here
as $4.25\times10^{22}$ $\mbox{cm}^{-3}$, because
$\lambda_{\mbox{tr}}(k_1^{(1)}\rightarrow 0) \sim const$.

\narrowtext
\section{Numerical Results}

We employ  
muonic atomic unit: distances are measured in units of $a_\mu$, where 
$a_\mu$ is the radius of muonic hydrogen atom.  
The integro-differential equations were solved  by discretizing  them into
a linear system of equations. The integrals in Eqs. (\ref{eq:most1})
and
(\ref{eq:most2}) are discretized using the  trapezoidal rule
and the partial derivatives are  discretized using a three-point rule
\cite{as}. 
The discretized equation is subsequently solved by Gauss
elimination method.
As we are concerned with the low-energy limit 
only  the total angular momentum $L=0$
is  taken into account. 
Even at zero incident energy, the transfer channels
are open and their wave functions are rapidly oscillating Coulomb waves. 
In order to get a converged solution we needed a large number of 
discretization points (up to 900) adequately distributed between 
0 to 40$a_\mu $. More points are taken near the origin
where the interaction potentials are large; a smaller number of points are 
needed at large distances.  For example, near the origin we took up to 40
equally spaced 
points per an  unit length interval $a_\mu$, in the intermediate region
($\rho = 10 - 20 a_\mu$) we took  up to 25 equally spaced
points per   unit length interval $a_\mu$, and in the asymptotic region 
($\rho = 20 - 40 a_\mu$) we took  up to 15 equally spaced
points per   unit length interval $a_\mu$. 
The following mass values are used in the unit
of electron mass:\hspace{1mm}
$m({^1{\mbox H}})$ = 1836.152, \hspace{1mm}
$m({^2{\mbox H}})$ = 3670.481, \hspace{1mm}
$m({^3{\mbox {He}}})$ = 5495.882,\hspace{1mm} 
$m({^4{\mbox {He}}})$ = 7294.295, \hspace{1mm}
$m({^6{\mbox {Li}}})$ = 10961.892,\hspace{1mm}
$m({^7{\mbox {Li}}})$ = 12786.385 and the muon mass is \hspace{1mm}
$m_{\mu}$ = 206.769.

We present muon-transfer rates
$\lambda_{\mbox{tr}}$ calculated using  the  formulation of last section
for processes (\ref{1}). 
First, we restrict ourselves to a two-level approximation by choosing
in the relevant close-coupling expansion  the hydrogen-like
ground states $(\mbox{H}_\mu)_{1s}$ and $({\mbox{X}^Z}_\mu)_{1s}$, where
H $={^1\mbox{H}}$ and $^2\mbox{H}$, and X$^Z$ = $^3{\mbox{He}}^{++}$, 
$^4{\mbox {He}}^{++}$, $^6{\mbox {Li}}^{+++}$ and $^7{\mbox {Li}}^{+++}$. 
Numerically stable and converged results were obtained in these cases. 
The rates $\lambda_{\mbox{tr}}$ $/10^6$ sec$^{-1}$
at low energies are presented in table 1
together with the results of \cite{Matv-1973,W-1992,Sultanov-1999}.
The results in this case converged to the precision shown in this table,
except in the case of 
$ {^2{\mbox H}_\mu} + {^4{\mbox {He}}^{++}} $, where it was difficult to
get converged result.
The present results are consistent with the experimentally observed
isotope effect \cite{iso1,Schell-1993,iso2}, e.g., the rate
decreases from $^1$H to $^2$H.

In table 2 we present our results for
transition rate of reaction (\ref{1}) to $(^3{\mbox {He}}^{++}_\mu)_{1s}$,
$(^6{\mbox {Li}}^{++}_\mu)_{1s}$ and 
$(^7{\mbox {Li}}^{++}_\mu)_{1s}$ from 
($^2$H$_{\mu}$)$_{1s}$ using the six-state close-coupling
model. The six states are H$_\mu$(1s,2s,2p) and X$^Z_\mu$(1s,2s,2p).
The results so obtained are consistent with the measured isotope
effect. The effect of including the (2s,2p) states in the calculational
scheme is also explicit there.  

The results reported in table 1 and 2
demonstrate the efficiency of  the present few-body model 
in describing muon transfer from H isotopes to  nuclei of
charge $Z_1 = 2$.
Its application to nuclei  involving higher charges,
therefore, is also expected to be  justified. The present 
calculation  with
$^6{\mbox {Li}}^{+++}$ or
$^7{\mbox {Li}}^{+++}$  represents the first examples
for such a full quantum-mechanical extension within the six-state
close-coupling model.

The study of three-body charge transfer reactions with Coulomb
repulsion in the final state has been the subject of this  work.
We have studied such reactions employing a 
detailed few-body description of the
rearrangement scattering problem by solving the Faddeev-Hahn-type
equations in coordinate space. To provide correct asymptotic
form in the final state the pure Coulomb interaction has
been incorporated directly into the equations. It is shown that within
this formalism, the application of a close-coupling-type ansatz
leads to satisfactory results already in low-order approximations
for direct muon-transfer reactions between hydrogen isotopes and light
nuclei ${\mbox {He}}^{++}$ and ${\mbox {Li}}^{+++}$. 
Because of computational difficulties, in this preliminary
application we have considered up to six states in the expansion
scheme (1s,2s,2p on each center $-$ (H$_\mu$) and X$_\mu^Z$), which
may not always be
adequate. Further calculations with larger basis sets are needed
to obtain accurate converged results.
However, the inclusion of three basis states on each center is expected to 
build in   a satisfactory account of the polarization potential in the
model. It has been observed \cite{psf} in studies of positron and
positronium 
scattering  
using close-coupling type approach that once the 1s,2s,2p states of 
positronium and target states are included, a good account of 
scattering including  
transfer reaction is obtained (estimated error of 
$10 - 20 \%$). However, the
inclusion of only the 1s basis functions do not lead to the converged
results. 
A similar conclusion can be obtained from tables 1 and 2. 
In view of the results of ref. \cite{psf}  we do not
believe the results of table 2 to be very different from the converged
ones, although we cannot provide a quantitative measure of convergence.
If the above conclusion based on the works of ref. \cite{psf} hold in this
case we expect a maximum error of $20 \%$ in table 2. 
  
Because of 
the present  promising results for the muon-transfer rates
of (\ref{1}) for $Z_1<4$, it seems useful to make
future applications of the
present formulation for larger targets with $Z_1 \ge 4$. Such
calculations involving nuclei of higher charge are in progress.
The present approach should also be useful in rearrangement 
collision involving electron, e.g., such as in 
H(1s) + He$^{++} \to$  H$^+$  +  He$^+$(1s),  considered in
\cite{Hose-1997}.


\acknowledgments
We acknowledge the support from FAPESP (Funda\c{c}\~{a}o
de Amparo \~{a} Pesquisa do Estado de S\~{a}o Paulo) of  Brazil.
The numerical calculations have been performed on the IBM SP2
Supercomputer of the Departamento de
F\'\i sica - IBILCE - UNESP,
S\~{a}o Jos\'e do Rio Preto, Brazil.


\mediumtext
\begin{table}
{Table 1.
Low energy muon transfer rates $\lambda_{\mbox{tr}}$$/10^6$
$\mbox{sec}^{-1}$
from proton  (${^1{\mbox H}_\mu}$)$_{1s}$
and deuteron (${^2{\mbox H}_\mu}$)$_{1s}$
to hydrogen-like ground state
(${^3{\mbox {He}}}^{+}_\mu$)$_{1s}$, 
(${^4{\mbox {He}}}^{+}_\mu$)$_{1s}$, 
(${^6{\mbox {Li}}}^{++}_\mu$)$_{1s}$ and
(${^7{\mbox {Li}}}^{++}_\mu$)$_{1s}$ 
within two-state close-coupling model.}
\vspace{4mm}
\begin{tabular}{lccccccccccc}
\multicolumn{1}{l}{System}                  &
\multicolumn{1}{c}{Energy}                  &
\multicolumn{1}{c}{Present Results}         &
\multicolumn{1}{c}{\ \ \cite{Sultanov-1999}}&
\multicolumn{1}{c}{\ \cite{W-1992}}         &
\multicolumn{1}{c}{\ \cite{Matv-1973}}\\
\multicolumn{1}{c}{ }                       &
\multicolumn{1}{c}{(eV)}                    &
\multicolumn{1}{c}{H$_{\mu}$(1s)$,$X$^Z_{\mu}$(1s)} &
\multicolumn{1}{c}{ }                       &
\multicolumn{1}{c}{ }                       &
\multicolumn{1}{c}{ }                       &
\\ \hline \hline
$ {^1{\mbox H}_\mu} + {^3{\mbox {He}}^{++}} $
& $\le 0.04 $ 
& $\dec 8.4 $ & $ \dec 7.25 $ & $\dec 10.9 $ & $\dec 6.3 $\\
& $\dec 0.1 $
& $\dec 8.3 $ & $ $ & $ $ & $ $ & $ $\\
& $\dec 1.0 $
& $\dec 8.1 $ & $ $ & $ $ & $ $ & $ $\\ \hline
$ {^1{\mbox H}_\mu} + {^4{\mbox {He}}^{++}} $
& $\le 0.04 $ 
& $\dec 6.8 $ & $ \dec 6.65 $ & $\dec 10.7 $ & $\dec 5.5 $
\\ \hline \hline
$ {^2{\mbox H}_\mu} + {^3{\mbox {He}}^{++}} $
& $\le 0.04 $
& $\dec 5.2 $ & $ \dec 4.77 $ & $\dec 9.6  $ & $\dec 1.3 $\\
& $\dec 0.1 $
& $\dec 5.1 $ & $ $ & $ $ & $ $\\
& $\dec 1.0 $
& $\dec 4.7 $  & $ $ & $ $ & $ $ & $ $\\ \hline
$ {^2{\mbox H}_\mu} + {^4{\mbox {He}}^{++}} $
& $\le 0.04 $
& $\hspace{-2mm} 5.0\pm 0.3 $ & $\dec 4.17 $ & $\dec 9.6  $ &
$\dec 1.0 $
\\ \hline \hline
$ {^2{\mbox H}_\mu} + {^6{\mbox {Li}}^{+++}} $
& $\le 0.04 $
& $\dec 1.2 $ & $ \dec 1.01 $ & & \\
& $\dec 0.1 $
& $\dec 1.2 $ & $ $ & $ $ & $ $\\
& $\dec 1.0 $
& $\dec 1.1 $  & $ $ & $ $ & $ $ & $ $\\ \hline
$ {^2{\mbox H}_\mu} + {^7{\mbox {Li}}^{+++}} $
& $\le 0.04 $
& $\dec 1.12 $ & $ \dec 0.96 $ &\\
& $\dec 0.1  $
& $\dec 1.12 $ & \\
& $\dec 1.0  $
& $\dec 1.06 $
\end{tabular}
\end{table}
\newpage

\begin{table}
{Table 2.
Low energy muon transfer rates $\lambda_{\mbox{tr}}$$/10^6$
$\mbox{sec}^{-1}$ from (${^2{\mbox H}_\mu}$)$_{1s}$ to 
hydrogen-like ground state
(${^3{\mbox {He}}}^{+}_\mu$)$_{1s}$, 
(${^6{\mbox {Li}}}^{++}_\mu$)$_{1s}$ and
(${^7{\mbox {Li}}}^{++}_\mu$)$_{1s}$ 
within six-state close-coupling model.}
\vspace{4mm}
\begin{tabular}{lccccccccccc}
\multicolumn{1}{l}{System}                  &
\multicolumn{1}{c}{Energy}                  &
\multicolumn{1}{c}{Present Results}         &\\
\multicolumn{1}{c}{ }                       &
\multicolumn{1}{c}{(eV)}                    &
\multicolumn{1}{c}{H$_{\mu}$(1s,2s,2p)$,$X$^Z_{\mu}$(1s,2s,2p)} &
\multicolumn{1}{c}{ }                       &
\multicolumn{1}{c}{ }                       &
\\ \hline \hline
$ {^2{\mbox H}_\mu} + {^3{\mbox {He}}^{++}} $
& $\le 0.04 $
& $\dec 9.0$\pm$ 0.2 $\\
& $\dec 0.1 $
& $\dec 8.8$\pm$ 0.2 $\\
& $\dec 1.0 $
& $\dec 5.0$\pm$ 0.2 $\\ \hline
$ {^2{\mbox H}_\mu} + {^6{\mbox {Li}}^{+++}} $
& $\le 0.04 $
& $\dec 1.9$\pm$ 0.1 $\\
& $\dec 0.1 $
& $\dec 1.9$\pm$ 0.1 $\\
& $\dec 1.0 $
& $\dec 1.2$\pm$ 0.1 $\\ \hline
$ {^2{\mbox H}_\mu} + {^7{\mbox {Li}}^{+++}} $
& $\le 0.04 $
& $\dec 1.6$\pm$ 0.1 $\\
& $\dec 0.1  $
& $\dec 1.6$\pm$ 0.1 $\\
& $\dec 1.0  $
& $\dec 1.2$\pm$ 0.1 $
\end{tabular}
\end{table}

\begin{thebibliography}{99}

\bibitem {Rafelski-1991}
Rafelski H E, Harley D, Shin G R and Rafelski J 1991
J. Phys. B: At. Mol. Opt. Phys. {\bf 24} 1469

\bibitem {Post-1999}
Hayaishi T, Tanaka T, Yoshii H, Murakami E, Shigemasa E,
Yagishita A, Koike F and Morioka Y 1999
J.Phys.B: At. Mol. Opt. Phys. {\bf 32} 1507

\bibitem {Jacot-1998}
Tresch S, Jacot-Guillarmod R, Mulhauser F, Schaller L A,
Schellenberg L, Schneuwly H, Thalmann Y -A and
Werthm\"uller A 1998 Euro. Phys. J. D {\bf 2} 93

\bibitem {Tresch1-1998}
Tresch S, Jacot-Guillarmod R, Mulhauser F, Piller C, Schaller L A,
Schellenberg L, Schneuwly H, Thalmann Y A,
Werthm\"uller A, Ackerbauer P, Breunlich W H, Cargnelli M, Gartner B,
King R, Lauss B, Marton J, Prymas W,
Zmeskal J, Petitjean C, Chatellard D, Egger J P, Jeannet E,
Hartmann F J and Muhlbauer M 1998 Phys. Rev. A {\bf 57} 2496

\bibitem {Wert-1998}
Wertm\"uller A,
Adamczak A, Jacot-Guillarmod R, Mulhauser F,
Schaller L A, Schellenberg L, 
Schneuwly H, Thalmann Y A, Trecsh S 1998
Hyperf. Interact. {\bf 116} 1

\bibitem{iso1}
Jacot-Guillarmod R, Mulhauser F, Piller C, Schaller L A,
Schellenberg L,
Schneuwly H, Thalmann Y A, Tresch S,
Werthm\"uller A  and Adamczak A 1997 Phys.  Rev.  A {\bf   55}  3447


\bibitem {Thalmann-1998}
Thalmann Y -A, Jacot-Guillarmod R, Mulhauser F, Schaller L A,
Schellenberg L, Schneuwly H, Tresch S and Wertm\"uller A 1998
Phys. Rev. A {\bf 57} 1713

\bibitem {Mulhauser-1993}
Mulhauser F and Schneuwly H 1993
J. Phys. B: At. Mol. Opt. Phys. {\bf 26} 4307

\bibitem {Schell-1993}
Schellenberg L 1993 Hyperf. Interact. {\bf 82} 513

\bibitem {Landau-1932}
Landau L D 1932 Z. Phys. Sow. Un. {\bf 2} 46;
Zener C 1932 Proc. Roy. Soc. A {\bf 137} 696

\bibitem{iso2}Bertin A, Bruno M, Vitale A, Placci A and Zavattini E
1973 Phys. Rev. A {\bf 7} 462

\bibitem{iso3} Haff P K, Rodrigo E and Tombrello T A 1977
Ann. Phys. (N.Y.) {\bf 104} 363

\bibitem {Schell-1996}
Schellenberg L, Adamczak A, Jacot-Guillarmod R, Mulhauser F,
Piller C, Schaller L A, Schneuwly H, Thalmann Y A, Trecsh S
and Wertm\"uller A 1996 Hyperf. Interact. {\bf 102} 215

\bibitem {Matv-1973}
Matveenko A V and Ponomarev L I 1972 Zh. Eksp. Teor. Fiz. {\bf 63} 48
(1973 Sov. Phys. JETP {\bf 36} 24)

\bibitem{W-1992}
Czaplinski W and Mikhailov A I 1992 Phys. Lett. A {\bf 169} 181

\bibitem {Sultanov-1999}
Sultanov R A, Sandhas W and Belyaev V B 1999 Euro. Phys. J. D {\bf 5} 33

\bibitem {Faddeev-1961}
Faddeev L D 1960 Zh. Eksp. Teor. Fiz. {\bf 39} 1459
(1961 Sov. Phys.$-$JETP {\bf 12} 1014)

\bibitem {Hahn-1968}
Hahn Y 1968 Phys. Rev. {\bf 169} 794
 
\bibitem {renat}
Sultanov R A 1999 Few Body Syst. Suppl. {\bf 10} 281

\bibitem {renat98}
Sultanov R A 1998 {\it Innovative Computational Methods in Nuclear
Many-Body Problems} ed H Horiuchi, Y  Fujiwara, M  Matsuo,
M  Kamimua, H  Toki  and Y  Sakuragi (Singapore: World Scinetific)
p 131

\bibitem {Hahn-1972}
Hahn Y and Watson K M 1972 Phys. Rev. A {\bf 5} 1718


\bibitem {Mott-1965}
Mott N F and H. S. W. Massey H S W 1965 {\it The theory of
atomic collisions} (London: Clarendon)

\bibitem{as}Abramowitz M and Stegun I A 1968 {\it Handbook of Mathematical
Functions}, (New York: Dover Publications), page 884, eq. (25.3.23), and 
page 885, eq, (25.4.1)

\bibitem{psf}
Mitroy J and Stelbovics A T 1994 J. Phys. B {\bf 27} 3257

Mitroy J and Stelbovics A T 1994 J. Phys. B {\bf 27} L79

Mitroy J and Stelbovics A T 1994 Phys. Rev. Lett.  {\bf 72} 3495

Chaudhuri P and Adhikari S K 1998 J. Phys. B {\bf 31} 3057

Chaudhuri P and Adhikari S K 1998 Phys. Rev. A  {\bf 57} 984


\bibitem {Hose-1997}
Hose G 1997 Phys. Rev. A {\bf 56} 1364

\end{thebibliography}
\end{document}